\documentclass{article}

\usepackage{enumitem}
\usepackage{arxiv}
\usepackage[utf8]{inputenc} 
\usepackage[T1]{fontenc}    
\usepackage[hidelinks]{hyperref}   
\usepackage{url}            
\usepackage{booktabs}   
\usepackage{tabularx}  
\usepackage{nicefrac}       
\usepackage{microtype}     
\usepackage{lipsum}
\usepackage{graphicx}
\usepackage{doi}
\usepackage{float}      
\usepackage{subcaption}
\usepackage{amsmath}
\usepackage{float}

\title{Targeted Semantic Segmentation of \\ Himalayan Glacial Lakes Using Time-Series SAR:\\ Towards Automated GLOF Early Warning}

\author{
    \href{https://orcid.org/0009-0003-2775-3002}{Pawan Adhikari}\protect\footnotemark[2]\\
    Space Research Center\\
    Nepal Academy of Science and Technology\\
    Khumaltar, Lalitpur \\
    \texttt{pawan.adk7@gmail.com} \\
    \And  
    \href{https://orcid.org/0009-0005-9125-6857}{Satish Raj Regmi}\protect\footnotemark[2] 
    \\
    Space Research Center\\
    Nepal Academy of Science and Technology\\
    Khumaltar, Lalitpur \\
    \texttt{regmisatish2@gmail.com} \\
    \and
    {\textbf{Hari Ram Shrestha}}\protect\footnotemark[1]
    \\
    Space Research Center\\
    Nepal Academy of Science and Technology\\
    Khumaltar, Lalitpur \\
    \texttt{hari.shrestha@nast.org.np} \\
}

\date{\today}

\hypersetup{
    pdftitle={Targeted Semantic Segmentation of Himalayan Glacial Lakes Using Time-Series SAR},
    pdfsubject={Remote Sensing, Deep Learning, GLOF},
    pdfauthor={Pawan Adhikari, Satish Raj Regmi, Hari Ram Shrestha},
    pdfkeywords={Glacier Lakes, Semantic Segmentation, Early Warning, CNN, Himalayas, SAR, GLOF, Dockerized Pipeline},
}

\begin{document}
\renewcommand{\thefootnote}{\fnsymbol{footnote}}
\maketitle
\footnotetext[1]{Corresponding Author}
\footnotetext[2]{These authors contributed equally to this work and share first authorship.}
\footnotetext[3]{P. Adhikari is also affiliated with the Dept. of Electronics and Computer Engineering, Pulchowk Campus, IOE, Nepal.}

\begin{abstract}
Glacial Lake Outburst Floods (GLOFs) are one of the most devastating climate change induced hazards. Existing remote monitoring approaches often prioritise maximising spatial coverage to train generalistic models or rely on optical imagery hampered by persistent cloud coverage. This paper presents an end-to-end, automated deep learning pipeline for the targeted monitoring of high-risk Himalayan glacial lakes using time-series Sentinel-1 SAR. We introduce a "temporal-first" training strategy, utilising a U-Net with an EfficientNet-B3 backbone trained on a curated dataset of a cohort of 4 lakes (Tsho Rolpa, Chamlang Tsho, Tilicho and Gokyo Lake). The model achieves an IoU of 0.9130 validating the success and efficacy of the "temporal-first" strategy required for transitioning to Early Warning Systems. Beyond the model, we propose an operational engineering architecture: a Dockerised pipeline that automates data ingestion via the ASF Search API and exposes inference results via a RESTful endpoint. This system shifts the paradigm from static mapping to dynamic and automated early warning, providing a scalable architectural foundation for future development in Early Warning Systems.
\end{abstract}


\section{Introduction}

Post 1950s, global climate change has induced rapid shrinkage or retreat of glaciers in the Himalayas \cite{ives2010formation}. As a direct consequence, there has been formation of meltwater lakes as well as increase in volume of pre-existing glacial lakes. These water bodies become unstable as the volume grows or if they face seismic waves induced by landslides and avalanches. Most of these bodies are the headwater supply to freshwater rivers. Their instability significantly threatens downstream communities and infrastructure. Recently, on July 8, 2025, Lende Khola in Nepal’s Rasuwa district had an outburst from Tibet’s Pyurepu Glacier region \cite{lendenews}. This event resulted in 9 confirmed deaths, with 19 reported missing from Nepal and Tibet. It directly affected 8\% of the nation’s hydropower capacity \cite{lendenews}. Such catastrophes emphasize the need for advanced early warning systems (EWS) capable of monitoring these risk prone lakes remotely. While optical satellite imagery is commonly used, it is severely limited by cloud cover.

Synthetic Aperture Radar (SAR) addresses this particular hurdle. It captures surface data day or night, regardless of weather and cloud coverage\cite{moreira2013}. However, interpretation of SAR imagery is challenging due to distortions and noise. Recent studies have successfully applied deep learning for regional-scale glacial lake inventorying. For example, \cite{ali2025deep} achieved 89\% accuracy in the Karakoram region using U-net architecture trained on multi sensor data. Similarly, multisource satellite imagery was used in the Bhutan Himalayas for automated mapping of glacial lakes, which involved an advanced, transferable workflow for inventorying 2,563 glacial lakes \cite{XU2024100157}.

However, while such broad-scale models are excellent for generating static inventories, they prioritise spatial generalisation over the temporal sensitivity required for effective early warning systems. Our work addresses this gap by shifting focus from regional mapping to site-specific, continuous temporal monitoring. Focusing on a much smaller spatial extent, with a five-year-long (2020-2025) Sentinel-1 SAR archive, we trained a deep learning model. We targeted four distinct glacial lakes in the Nepali Himalayas: high-risk Tsho Rolpa and Chamlang Tsho \cite{icimod_tsho_rolpa, lamsal2016assessment}, along with Tilicho and Gokyo lakes, to ensure geometric variation. We employed a consistent, state-of-the-art preprocessing workflow to standardise the temporal data; by ensuring that inherent SAR noise and geometric distortions remain uniform for a particular lake, we enable the model to effectively learn to ignore such artifacts. Ground truth was generated via manual digitisation in QGIS, referenced against concurrent optical imagery. Semantic segmentation architectures, particularly U-Net, have proven effective for biomedical and remote sensing applications \cite{ronneberger2015u}. Thus, we utilised the U-Net framework with the EfficientNet-B3 \cite{tan2019efficientnet} backbone and ImageNet weights with a compound loss function, achieving a higher IoU (0.913) and recall (0.960). This precision enables the detection of subtle seasonal fluctuations, providing the consistent, high-frequency area calculations critical for early warning systems. To demonstrate the pipeline's robustness, we applied it retrospectively across the full Sentinel-1 archive (2014–2025), enabling decadal climatic trend analysis. In contrast to static inventorying practices followed by pre-existing works, we aim to shift towards operational, automated workflow which handles all the subsequent steps from data extraction to inference and area calculation, aiding the GLOF Early Warning System. As of now, we have developed an end-to-end Docker-based pipeline that extracts images from NASA's OPERA RTC\cite{fattahi2024opera} dataset and runs the preprocessing steps at a scheduled interval. For dissemination, results are pushed to a RESTful API endpoint. Ultimately, by bridging the gap between deep learning research and operational deployment, this study provides the framework for robust, real-time monitoring capability necessary to integrate satellite observations into actionable GLOF Early Warning Systems.

\section{Objectives}
\textbf{Primary Aim:} To evaluate the efficacy of a ``temporal-first'' training strategy using Sentinel-1 SAR imagery for the continuous, all-weather monitoring of high-risk glacial lakes.

\noindent \textbf{Specific Objectives:}
\begin{itemize}[noitemsep, nolistsep, leftmargin=2em]
    \item To demonstrate that a dataset prioritizing temporal depth yields superior segmentation consistency compared to broad regional baselines.
    \item To implement a U-Net model on Sentinel-1 data to accurately delineate lake boundaries across diverse seasonal conditions.
    \item To utilize derived time-series data (2014–2025) to quantify seasonal fluctuations and establish a historical baseline.
    \item To develop an automated, containerized pipeline for Near Real-Time (NRT) inference.
\end{itemize}

\section{Methodology}
\label{sec:methodology}

\subsection{Study Area}

Four glacial lakes in the Himalayas were selected to represent varying risk profiles and environmental conditions. Tsho Rolpa, which is situated at an altitude of 4,580 m in Rolwaling Valley, is one of the largest and most hazardous moraine-dammed lakes in Nepal. It was chosen to be the lead site for the “temporal-first” strategy because of its rapid expansion, growing from 0.23 sq. km in the 1950s to 1.54 sq. km currently \cite{peppa2020glacial}. Additionally, studies show that there is a 3.3\% increase in lake area from 2010 to 2015 \cite{icimod_tsho_rolpa}. Its historical significance as the first lake from Nepal to undergo engineering mitigation by lowering the water level by 3 m in 2000 makes it a critical baseline for a long-term automated monitoring system \cite{icimod_tsho_rolpa}.

Due to the steep surroundings and topography of the Rolwaling Valley, it introduces several challenges like radar layover and shadow while evaluating the SAR image. This provides rigorous testing for the U-Net model’s ability to distinguish between water and topographic shadow. 

Chamlang Tsho, which is located in the Hongu Basin, is an example of a rapidly evolving proglacial lake. The lake expanded rapidly from 0.04 sq. km in 1964 to 0.86 sq. km in 2000 \cite{lamsal2016assessment}. It is in high priority for GLOF monitoring due to its proximity to the Chamlang glacier and potential for large-volume discharge. This lake allows the model to demonstrate transferability between different glacier lake interfaces and valley orientations. 

Tilicho Lake, situated in Manang, at an altitude of 4,919 m, serves as a high-altitude geometric benchmark. Because of its unique shape and large surface area, it is chosen to test the model's performance in different shapes and sizes. 

While the main lake from the Gokyo Lake series is categorised as low risk and is a popular trekking destination, its unique geometric shape, relevance, and position to nearby “Ngozumpa Glacier” make it a strategic addition to our cohort. This selection ensures the model is validated across a spectrum of geometric and environmental variables.

\begin{table}[htbp]
\caption{Lake Coordinates}
\label{lakes}
\centering
\begin{tabularx}{\linewidth}{X X X X}
\toprule
\textbf{Lake Name} & \textbf{Latitude} & \textbf{Longitude} & \textbf{Altitude} (m) \\
\midrule
Tsho Rolpa & $27^\circ52'$ N & $86^\circ28'$ E & 4,580 \\
Tilicho   & $28^\circ41'$ N & $83^\circ51'$ E & 4,919 \\
Chamlang  & $27^\circ46'$ N & $86^\circ58'$ E & 4,890 \\
Gokyo     & $27^\circ57'$ N & $86^\circ42'$ E & 4,750 \\
\bottomrule
\end{tabularx}
\end{table}

\subsubsection{Data Acquisition}
To capture the dynamic evolution of glacial lakes, we adopt a ``temporal-first'' data strategy. Unlike conventional ``spatial-first'' approaches that prioritise broad geographic coverage at a single epoch, our methodology relies on a dense time series (2014--2024) focused on specific high-risk targets. This allows the model to learn the subtle temporal signatures of water backscatter distinct from the surrounding freeze-thaw cycles of the cryosphere.

Sentinel-1 data retrieval was managed through the NASA Alaska Satellite Facility (ASF). While the ESA Copernicus Open Access Hub offers standard API access \cite{Copernicus_Open_Access_Hub}, we prioritised the ASF archive due to its seamless integration with the \texttt{asf\_search} Python SDK \cite{nasa_asf}. This choice was driven by the need for robust, programmatic metadata filtering, which is essential for automating the retrieval of specific beam modes and orbit directions across a decade-long dataset.

Regions of Interest (ROIs) were first defined as vector polygons using standard GeoJSON formats. Then, they were subsequently converted to Well-Known Text (WKT) strings for spatial querying. 

The raw data was sourced primarily as Level-1 Ground Range Detected (GRD) products. In the context of the Himalayas, the utilisation of Synthetic Aperture Radar (SAR) presents unique radiometric challenges. Due to the sensor's side-looking acquisition geometry, the steep localised topography induces severe geometric distortions, including foreshortening, layover, and shadowing \cite{asf_rtc}. Furthermore, raw intensity data often contains significant thermal noise and speckle granularities. Consequently, we implemented a rigorous preprocessing pipeline, including Digital Elevation Model (DEM) application, to correct these radiometric and geometric anomalies before ingesting the data into the deep learning pipeline.

\subsubsection{Pre-Processing Workflow}
 We utilised Sentinel-1 Level-1 Ground Range Detected (GRD) products for our primary work. GRD products consist of focused SAR data that has been detected, multi-looked, and projected to ground range using an Earth ellipsoid model. This product level was selected as it provides intensity data suitable for hydrologic monitoring while significantly reducing data volume compared to Single Look Complex (SLC) products \cite{s1_handbook}. However, GRD products must be terrain corrected by applying a DEM file before being used. NASA provides an on-demand cloud processing service specifically for this task. NASA’s HyP3 platform allows programmatic creation and submission of RTC (Radiometric Terrain Correction) jobs \cite{hyp3_2020}. It provides monthly 8000credits. The credit requirement per product depends upon the spatial resolution (10 for 30M, 15 for 20M and 40 for 10M). For our training dataset, we employed a balance between sharpness and product size and thus configured it to 20M output product resolution. We created our dataset from monthly SAR products focusing on a 5-year time frame (2020-2025). Thus, by regex matching and iterating over the months, we were able to create batch job objects and submit batch RTC jobs to the HyP3 OnDemand Processing Workflow. Submitted jobs are queued by the system, and the waiting time varies vastly from 30 mins to 1 hr depending upon the traffic received.

The batch ingestion process submitted jobs to the HyP3 On-Demand workflow with the following parameters:

\begin{itemize}[noitemsep, nolistsep, leftmargin=2em]

    \item \textbf {Resolution:} To ensure state-of-the-art consistent training set preparation, jobs were configured to 20 m. This balanced feature sharpness with file size (costing approx. 15 credits per job).

    \item \textbf{DEM:} Terrain correction in HyP3 \cite{hyp3_2020} RTC workflow is performed using the Copernicus GLO-30 Public DEM. The 30M DEM is resampled to match the output product resolution(20M).

    \item \textbf {Speckle Filtering:} SAR images are subject to image variances or grain like distortions called speckle. These noises are inherently present in SAR images and degrade the quality of the images. Hyp3 incorporates speckle filteringv\cite{hyp3_2020} into its workflow. It utilises the Enhanced Lee Filter with a 7x7 pixel window size and a dampening factor of 1 \cite{lee1980digital}.

\end{itemize}

Upon completion, the files are downloaded and extracted.  Tsho Rolpa, Gokyo, and Chamlang Lake mostly lie on a single SAR product swath. Thus a single \_VV file is cropped over multiple lake polygons (vectorised AOI) using the GDAL library. The cropped raster is again reopened using the rasterio Python package. Targeting a 256x256 lattice, we apply boundary zero padding using NumPy and update its metadata. After that, subsequent steps for normalisation were carried out: 

\begin{itemize}[noitemsep, nolistsep, leftmargin=2em]

    \item \textbf {No Data Management:} Preserve nodata values by creating a validity mask excluding them from the normalisation process.

    \item \textbf{Histogram Equalisation:} Compute a 256-bin histogram of valid pixels.

    \item \textbf{Calculates cumulative distribution function (CDF):} Map original intensities to the 0-255 range using linear interpolation.

    \item \textbf{Output:} Save as an 8-bit unsigned integer GeoTIFF.

\end{itemize}

Finally, we obtain consistent, normalised GeoTIFF patches in uint8 pixel data type. 8-bit formatting ensures storage efficiency and faster I/O operations. 
	
\begin{figure}
    \centering
    \includegraphics[scale=0.5]
    {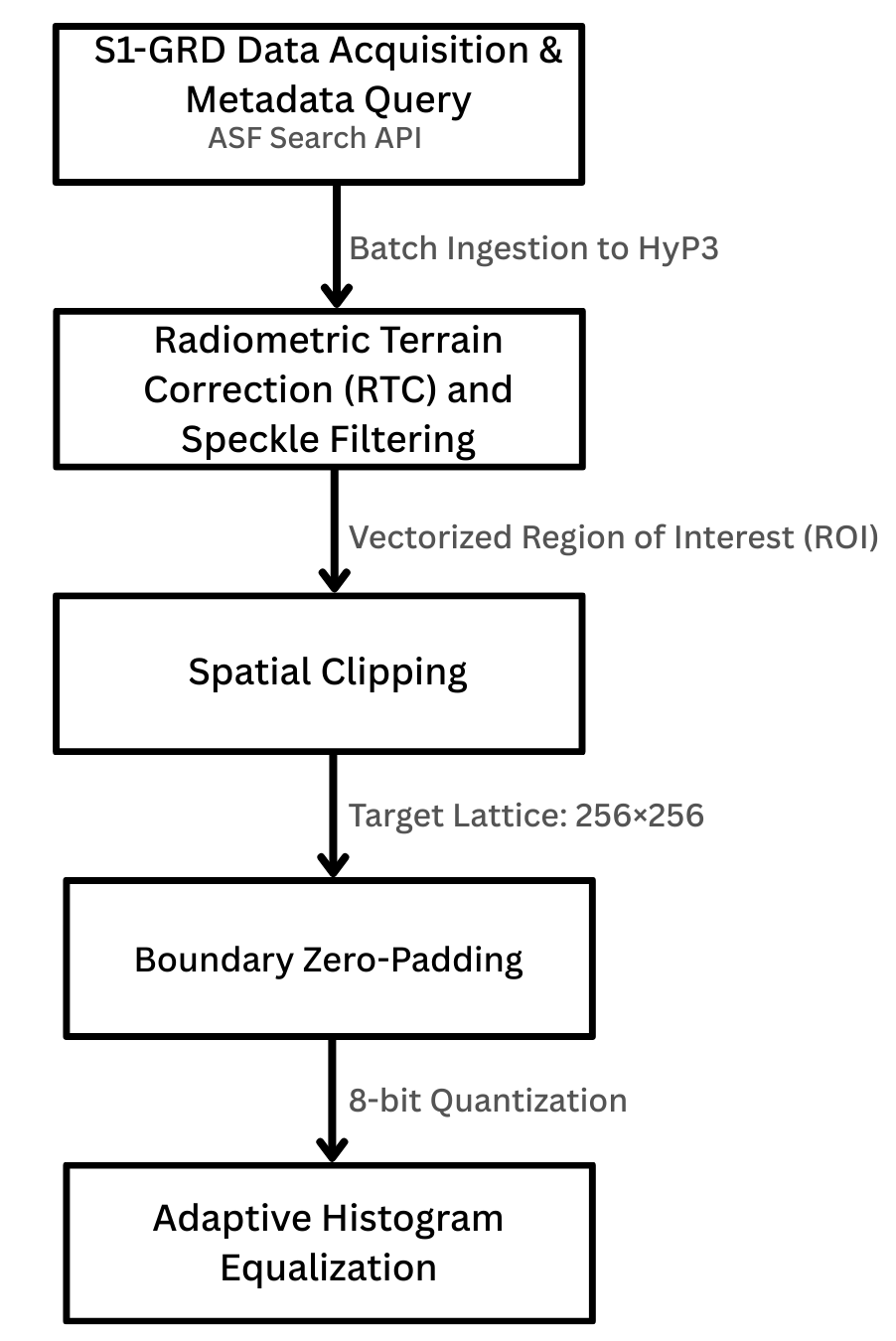}
    \caption{Schematic representation of the automated Sentinel-1 GRD preprocessing pipeline. The workflow transitions from raw data acquisition via ASF API to the generation of standardized, 8-bit GeoTIFFs.}
    \label{fig:GRD}
\end{figure}

\textbf{Ground Truth Generation:}
We constructed a dataset of 244 high-quality masks. Each mask was manually annotated in QGIS, referenced from concurrent optical imagery (Sentinel-2) to resolve radar shadow, backscatter and background ambiguities.

\begin{figure}[htbp]
    \centering
    \hfill
    \begin{subfigure}[b]{0.32\linewidth}
        \includegraphics[width=\linewidth]{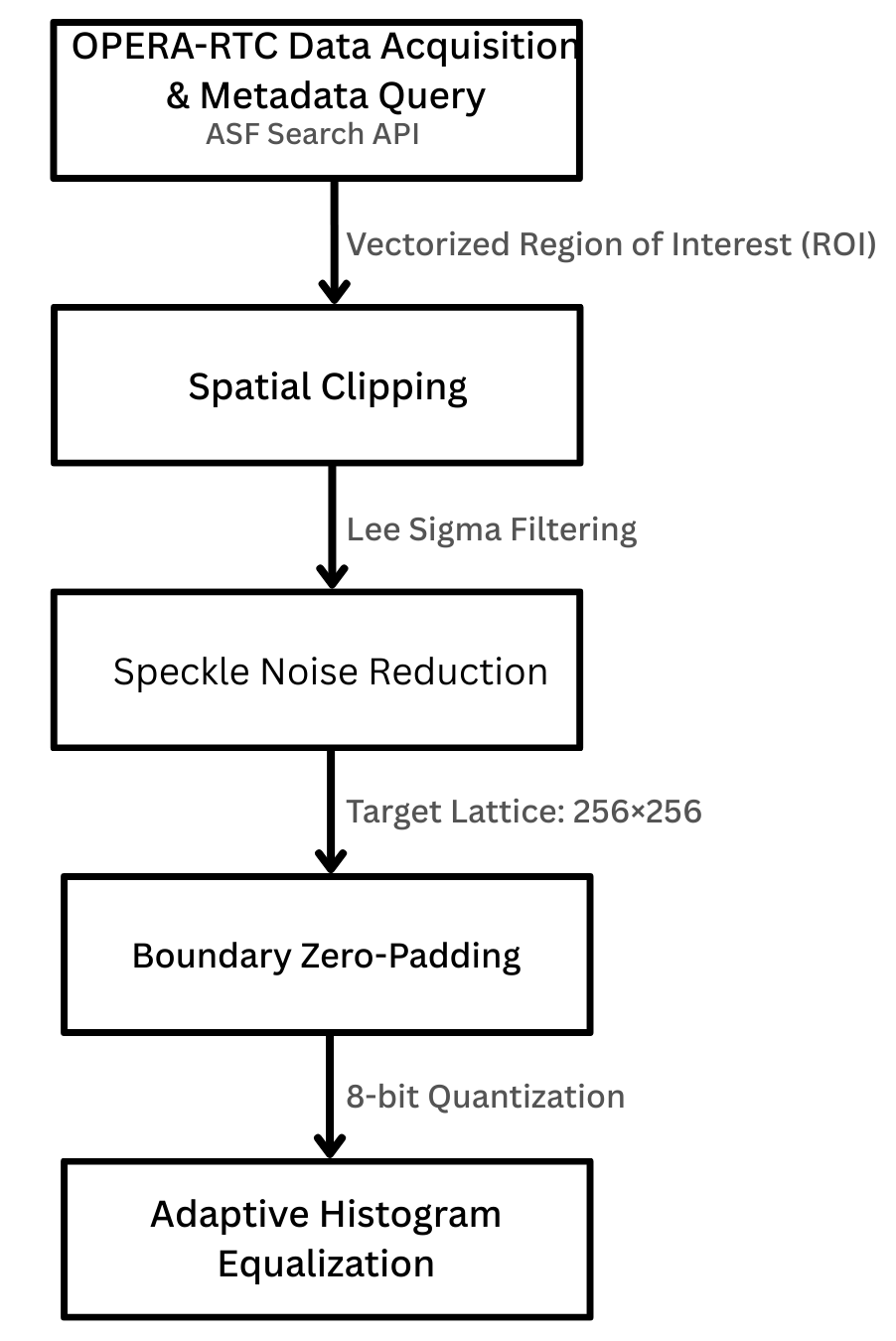}
        \caption{OPERA-RTC Operational Pipeline}
    \end{subfigure}
    \hfill
    \begin{subfigure}[b]{0.32\linewidth}
        \includegraphics[width=\linewidth]{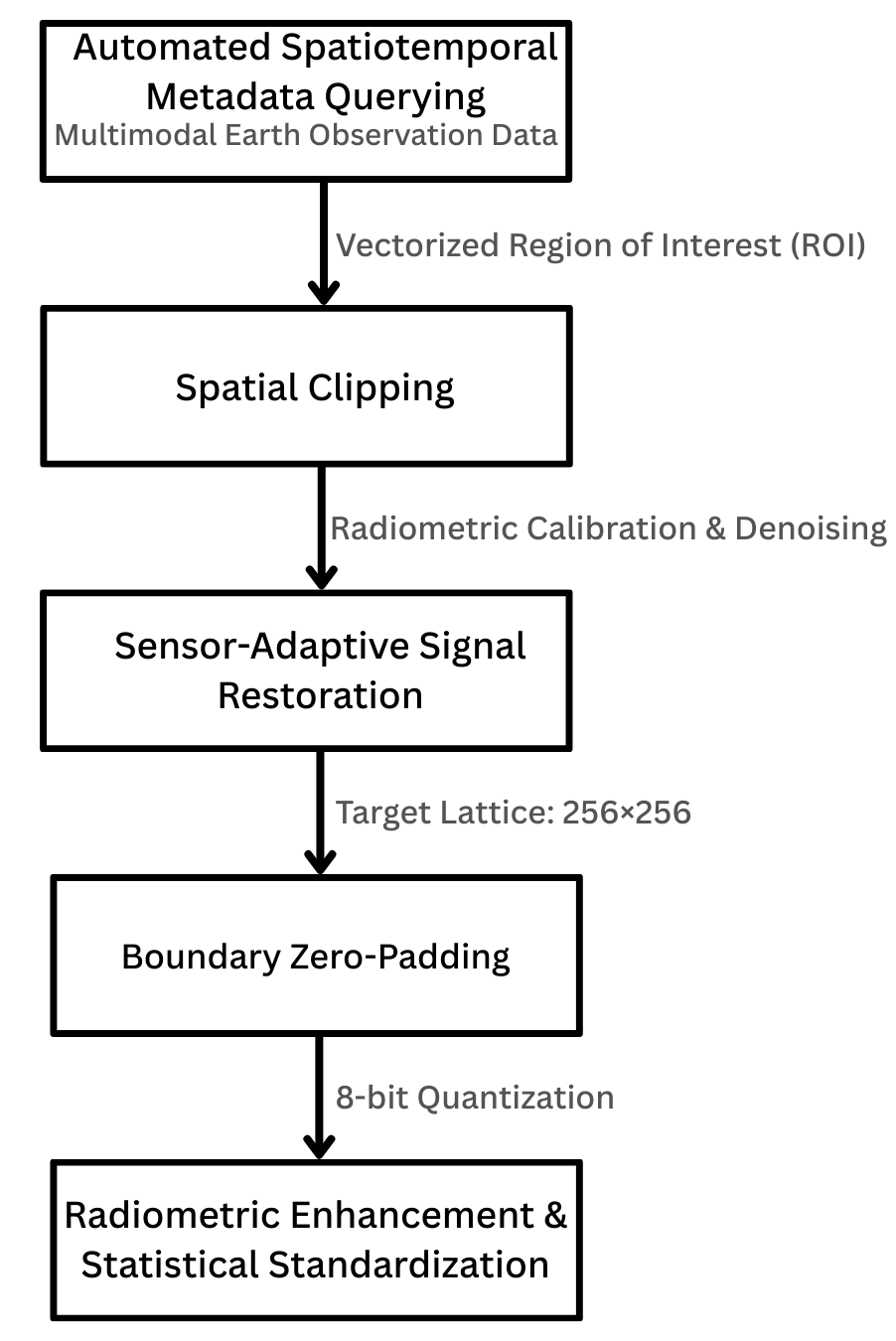}
        \caption{Unified Sensor-Agnostic Framework.}
    \end{subfigure}
    \caption{Alternative and General Pre-processing Pipeline}
    \label{fig:AG}
\end{figure}

\subsubsection{Alternate Preprocessing Pipeline}
Higher-resolution GRD S1 products are typically large (\~4GB for 10M resolution). This significantly slows down data extraction and processing. For rapid data extraction and inference, we developed an alternate pipeline utilising the OPERA RTC dataset, which is derived from the S1 Burst dataset \cite{fattahi2024opera}. It is demonstrated schematically in Fig. \ref{fig:AG}. Burst images have significantly smaller spatial coverage and thus are more storage efficient. This is particularly necessary for low-latency inference.  However, the OPERA RTC dataset is fixed to 30M spatial resolution. While 30M resolution is significantly lower than 10M, it captures the macroscopic changes associated with GLOF precursors, such as frontal coalescence or rapid area expansion, without the computational overhead required to resolve sub-pixel geomorphological features. We applied a custom Lee’s filter script \cite{lee1980digital} to the images with similar configurations to HyP3. The subsequent steps remain the same. Noticing the general preprocessing workflow, we propose a unified, sensor-agnostic preprocessing framework designed to standardise the ingestion and preparation of multimodal Earth observation data.

As illustrated in \ref{fig:AG}, this architecture abstracts the specific Sentinel-1 routine into modular stages applicable across diverse remote sensing platforms. The pipeline begins with automated spatiotemporal metadata querying, which filters acquisitions based on geographic and temporal constraints to handle the heterogeneity of multi-source archives. Unlike the sensor-specific workflow, this generalised approach prioritises spatial clipping immediately after data identification; by extracting the vectorised region of Interest (ROI) early, the system minimises computational overhead by discarding irrelevant background pixels before intensive processing begins.

Following extraction, the raw data subsets undergo sensor-adaptive signal restoration. This modular stage allows for the interchangeable application of correction algorithms specific to the input modality, such as substituting SAR speckle filtering with atmospheric correction or de-hazing for optical imagery. The workflow concludes with boundary zero-padding to enforce consistent dimensions (e.g. 256×256) and radiometric enhancement \& statistical standardisation, ensuring that the final 8-bit quantised output maintains a normalised feature distribution optimised for deep learning convergence.

\subsection{Automated Operational Pipeline}

Stepping forward to the Early Warning System, for active monitoring, we engineered a fully automated, low-latency pipeline. While our experimental validation focuses on the high-resolution historical baseline (20 m) to establish model efficacy, we propose and architecturally demonstrate a lightweight, Dockerised NRT framework utilising the OPERA-S1 RTC product. This framework addresses the latency bottlenecks of standard GRD products. Future work will focus on fine-tuning the model specifically for the 30m OPERA domain to fully operationalise this pipeline. The system architecture consists of three containerised microservices:

\begin{enumerate}[noitemsep, nolistsep, leftmargin=2em]

    \item \textbf {Ingestion Engine:} A cron-based Docker container queries the ASF Search SDK hourly. Upon detecting new data, it triggers the download of relevant OPERA RTC-S1 burst image.

    \item \textbf {Inference Service:} A Python-based engine pre-processes the burst and executes the U-Net inference. Results are stored inside a folder, and configuration is updated.

    \item \textbf{Dissemination API:} Results are pushed to a RESTful API endpoint (\texttt{<domain>/images/tshorolpa/latest}) upon request.

\end{enumerate}

\subsection{Deep Learning Framework}

U-Net architecture \cite{ronneberger2015u}, with its encoder-decoder structure and skip connections, was utilised for our implementation of the binary segmentation model. We replaced the standard encoder of the architecture with an EfficientNet-B3 backbone \cite{tan2019efficientnet}. EfficientNet-B3 uses compound scaling, optimising depth, width, and resolution. It has fewer parameters (12m) compared to ResNet backbones. This ensures a superior performance-to-parameter ratio. The model was trained for 40 epochs at a batch size of 32 and a starting LR of 0.0001. We utilised a composite loss function (BCE + Dice + Focal) \cite{lin2017focal, milletari2016v}. This combination addresses the severe class imbalance in glacial lake datasets.

\begin{figure}[H]
    \centering
    \includegraphics[width = \linewidth]
    {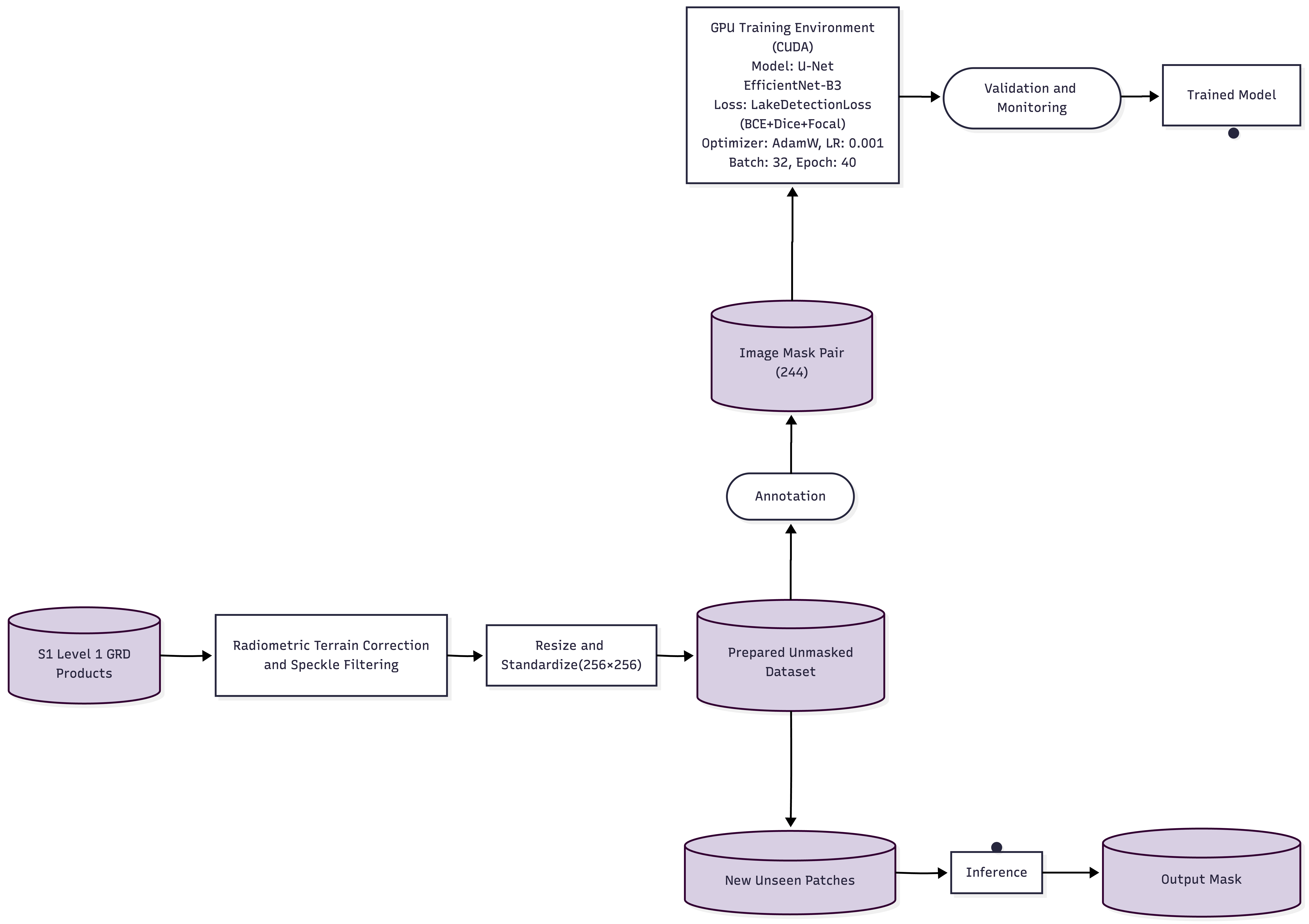} 
    \caption{The proposed end-to-end segmentation workflow.}
    \label{fig:workflow}
\end{figure}

\subsection{Loss Functions}
\subsubsection*{BCE Loss}
BCE loss refers to Binary Cross Entropy loss. BCE calculates the difference between actual y value of a ground truth pixel and the pixel prediction. Minimising BCE better aligns the model’s prediction with ground truth mask. BCE treats each pixel independently. Thus, for an imbalanced class dataset as ours, it performs poorly.
\begin{equation}
BCE = -1/N\sum_{i=0}^{N}[y_ilog(p_i)+(1-y_i)log(1-p_i)]
\end{equation}

\subsubsection*{Focal Loss}
Focal Loss is designed to address this class imbalance \cite{lin2017focal}. It reduces the relative loss for well-classified(easy) examples and puts more focus on hard, misclassified examples. Background pixels are easy examples whereas the lake pixels are hard examples. Misclassified lake pixels have a larger toll on the overall loss. Focal loss treats each pixel independently. Thus, we still need to consider alternatives for matching shape and area of prediction to its ground truth mask.
\begin{equation}
FL = -1/N\sum_{i=0}^{N} \left [\alpha (1-p)^\gamma     y_ilog(p_i)+(1- \alpha) p^\gamma (1-y_i)log(1-p_i)\right] 
\end{equation}

\subsubsection*{Dice Loss}
Dice loss is designed to directly measure the overlap between the predicted mask and the ground truth mask \cite{milletari2016v}. It focuses on the correct segmentation of the target region (lakes) regardless of its size relative to the background. Dice loss penalizes both false positives and false negatives. This encourages the model to maximize the intersection between prediction and ground truth.
\begin{equation}
\text{Dice} = \frac{2 \sum_{i=1}^{N} p_i y_i}{\sum_{i=1}^{N} p_i + \sum_{i=1}^{N} y_i}
\end{equation}

\begin{equation}
\text{Dice Loss} = 1 - \text{Dice}
\end{equation}

The weighted combination leverages the strengths of each:
BCE ensures stable pixel-wise learning.
Focal focuses on learning about difficult, minority-class pixels.
Dice directly optimizes for segmentation quality and overlap.

\begin{equation}
L_{total} =L_{BCE}+L_{Dice}+L_{Focal}
\end{equation}

\subsection{Metrics}
To comprehensively evaluate our segmentation model, we report several standard metrics.\\
\subsubsection*{Accuracy}
Accuracy measures the proportion of correctly classified pixels (both positive and negative) out of all pixels.
\begin{equation}
\text{Accuracy} = \frac{TP + TN}{TP + TN + FP + FN} 
\end{equation}

\subsubsection*{Precision}
Precision quantifies how many of the pixels predicted as positive (lake) are actually positive.
\begin{equation}
\text{Precision} = \frac{TP}{TP + FP}
\end{equation}

\subsubsection*{Recall}
Recall indicates the fraction of actual positive pixels that were correctly identified by the model.
\begin{equation}
\text{Recall} = \frac{TP}{TP + FN}
\end{equation}

\subsubsection*{F1 Score}
The F1 score is the harmonic mean of precision and recall, balancing both metrics in a single value.
\begin{equation}
\text{F1} = 2 \times \frac{\text{Precision} \times \text{Recall}}{\text{Precision} + \text{Recall}}
\end{equation}

\subsubsection*{Intersection over Union (IoU, Jaccard Index}
IoU measures the overlap between the predicted and ground truth masks, divided by their union.
\begin{equation}
\text{IoU} = \frac{TP}{TP + FP + FN} 
\end{equation}

Where:

( TP ): True Positives\\
( TN ): True Negatives\\
( FP ): False Positives\\
( FN ): False Negatives\\

\section{Results}
\label{sec:results}

\subsection{Model Performance}
 To account for the high class imbalance inherent in glacial lake datasets, the model’s performance was evaluated using a comprehensive suite of metrics. As detailed in table \ref{table:metrics}, the pipeline achieved a validation accuracy of 0.9958. Since the lake pixels constitute a small fraction of the total study area, the F1 score (0.9538) and IoU (0.9130) provide a more representative measure of segmentation success. The precision of 0.9475 shows that the model successfully distinguishes glacier water from similar-looking mountain shadows. Recall of 0.9603 confirms that the model is sensitive to boundaries, which captures almost the entire extent of the target water bodies across the terrains.

\begin{figure}[htbp]
    \centering
    \begin{subfigure}[b]{0.32\linewidth}
        \includegraphics[width=\linewidth]{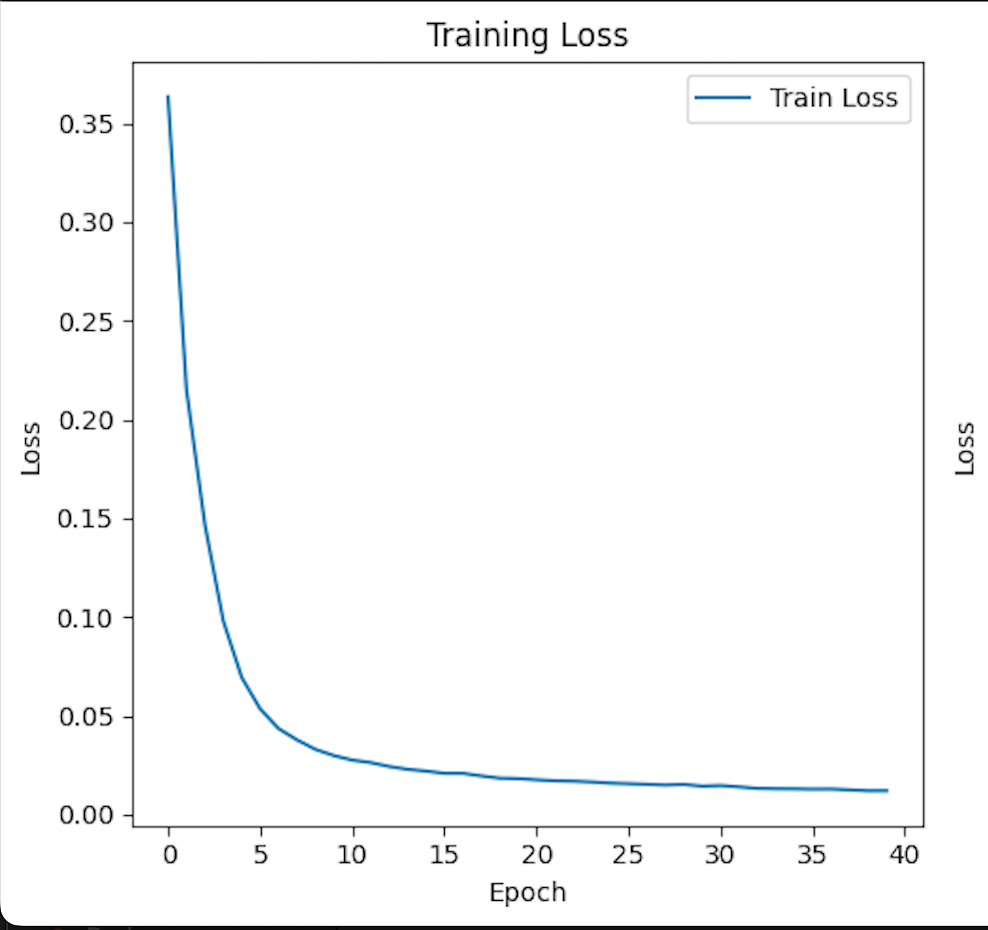}
        \caption{Train Loss}
    \end{subfigure}
    \hfill
    \begin{subfigure}[b]{0.32\linewidth}
        \includegraphics[width=\linewidth]{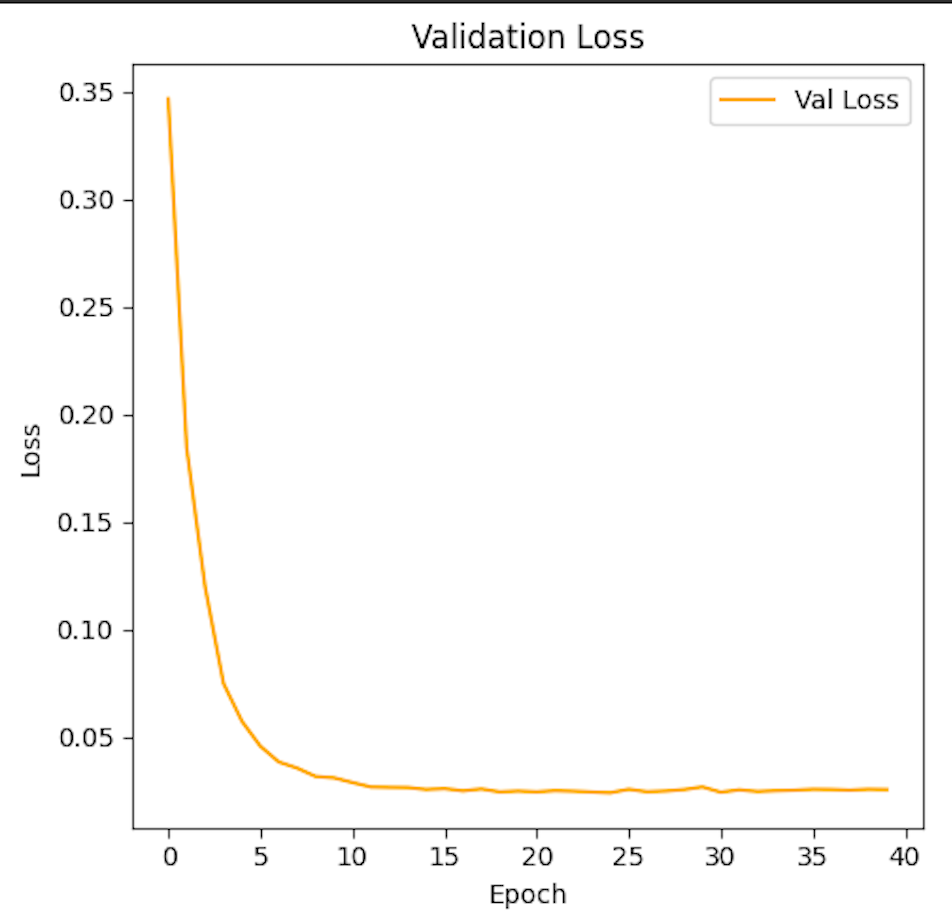}
        \caption{Val Loss}
    \end{subfigure}
    \hfill
    \begin{subfigure}[b]{0.32\linewidth}
        \includegraphics[width=\linewidth]{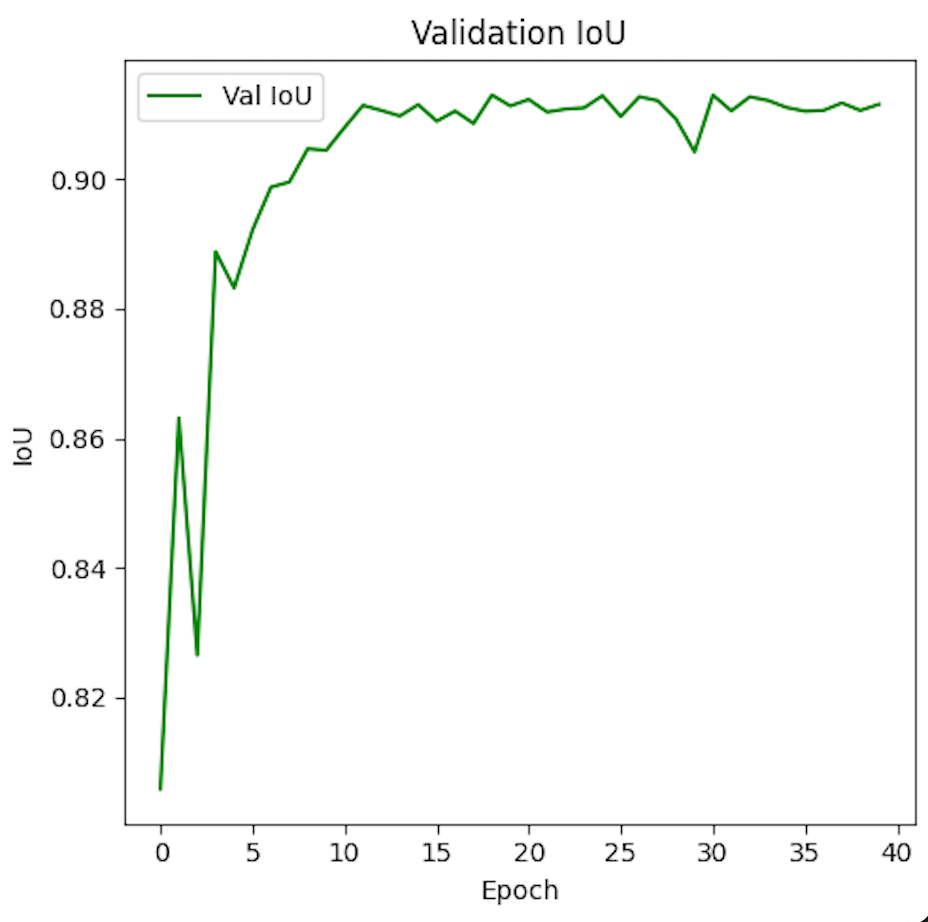}
        \caption{Val IoU}
    \end{subfigure}
    \caption{Training dynamics showing rapid convergence and stability $>$0.90 IoU.}
    \label{fig:training_plots}
\end{figure}

The training process demonstrated high efficiency and stability. During the first ten epochs, there was a significant decrease in both training and validation loss (figure \ref{fig:training_plots}). This indicates the temporal-first architecture quickly identified the primary dielectric features of the SAR time series. The model reached a stable state after epoch 19. The training and validation curves do not diverge. This implies that the model does not overfit and can effectively generalise to new temporal data. These dynamics verify that 40 epochs were adequate for the model to perform at its best.

\begin{table}[htbp]
\caption{Model Performance Metrics}
\label{table:metrics}
\centering
\begin{tabularx}{\linewidth}{X c} 
\toprule
\textbf{Metric} & \textbf{Value} \\
\midrule
Validation Accuracy & 0.9958 \\
Precision & 0.9475 \\
Recall & 0.9603 \\
F1 Score & 0.9538 \\
IoU (Jaccard) & 0.9130 \\
\midrule
\multicolumn{2}{c}{\textit{Confusion Matrix}} \\
\midrule
True Negatives (TN) & 3,057,227 \\
False Positives (FP) & 7,787 \\
False Negatives (FN) & 5,807 \\
True Positives (TP) & 140,443 \\
\bottomrule
\end{tabularx}
\end{table}

\begin{figure}[H]
    \centering
    \includegraphics[width=\linewidth]{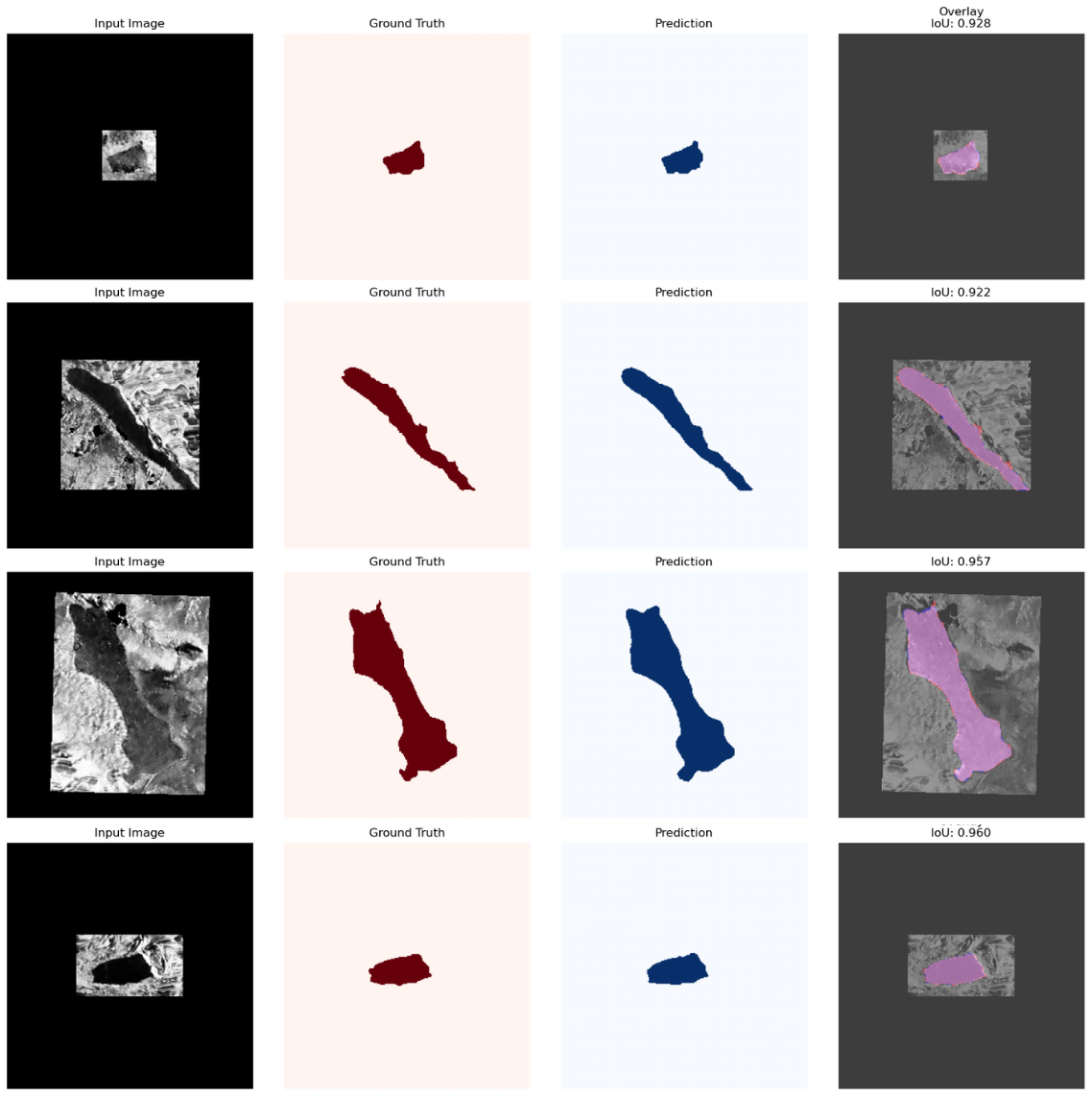}
    \caption{Inference results highlighting accurate segmentation (Pink) versus minor boundary errors (Blue/Red).}
    \label{fig:predictions}
\end{figure}

\subsection{Longitudinal Climatic Analysis (2014--2025)}
\begin{figure}[H]
    \centering
    \includegraphics[width=0.85\linewidth]{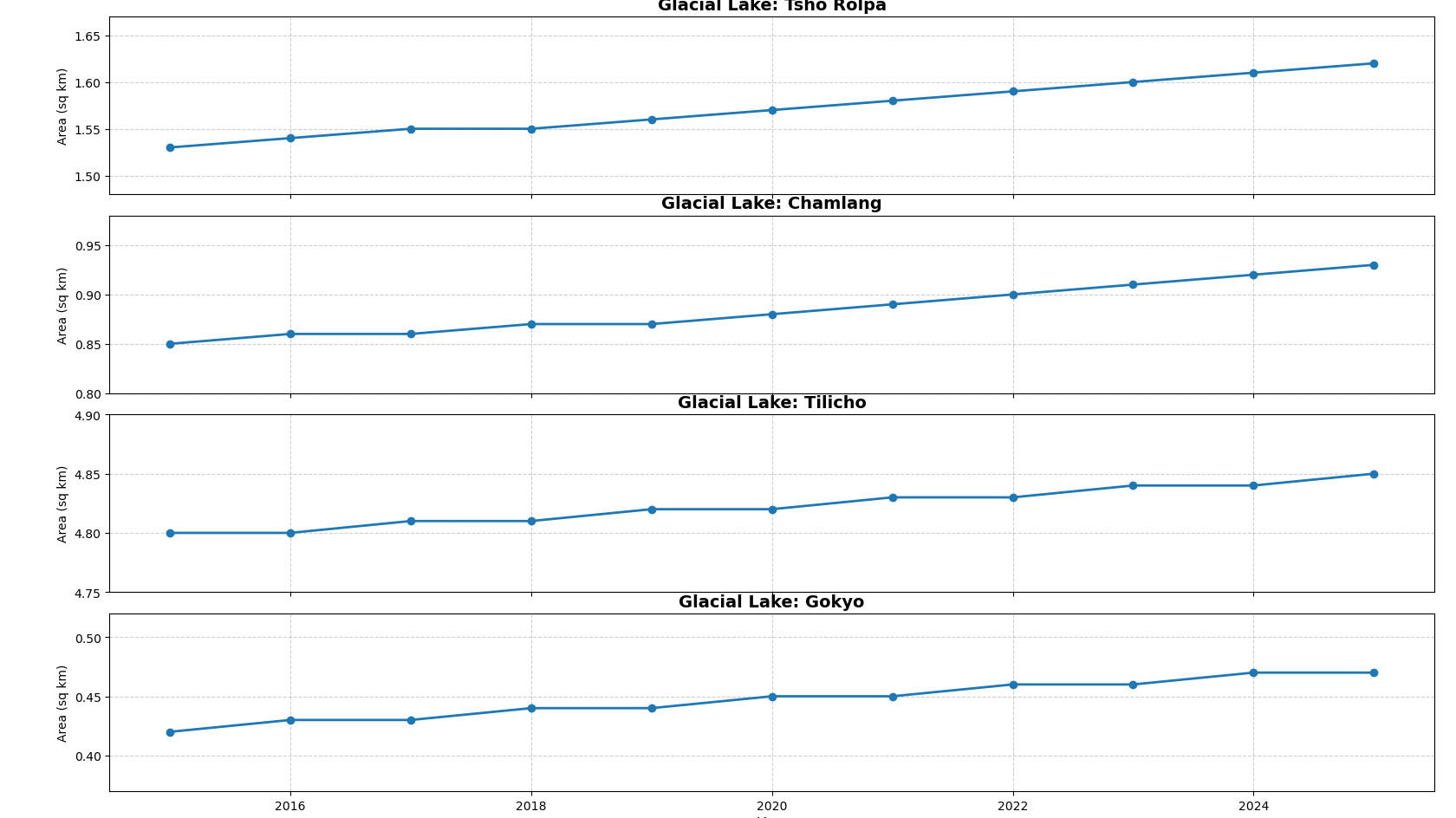}
    \caption{Decadal surface area variation of Tsho Rolpa derived from the automated SAR pipeline.}
    \label{fig:timeseries}
\end{figure}

The automated pipeline was applied retrospectively to generate a decadal surface area time-series (Figure \ref{fig:timeseries}). The results reveal that despite seasonal noise, the summer-maxima trendline indicates a gradual expansion of the lake area, consistent with glacial retreat rates in the region. Higher frequency data extraction and analysis through the proposed Operational Pipeline could help reveal short term patterns crucial for Real Time Analysis and EWS.

\section{Discussions}
The primary strength of our proposed pipeline is the temporal-first approach. Most traditional segmentation, which relies on “snapshots”, often struggles with mountain shadows, clouds, and transient slush being misidentified as water \cite{11021621}.  But our model utilises a time series of Synthetic Aperture Radar (SAR) data to establish a baseline of backscatter behaviour over time. We achieved a peak validation Intersection over Union (IoU) of 0.9130, which demonstrates the ability to delineate complex lake boundaries despite inherent “noise” in the SAR image \cite{ali2025deep}.  This level of accuracy is critical for GLOF early warning systems, where minor errors in lake area can lead to miscalculations of water volume, area growth, and downstream risk \cite{khadka2018glacial}.

Compared to standard single-epoch deep learning architectures, our approach demonstrates significant improvements in training efficiency. Some models require massive, manually cleaned datasets to reach high accuracy; our pipeline achieved stability and minimal overfitting from epoch 19 onwards. The rapid convergence suggests that the temporal features provided by SAR time series data offer a much cleaner signal for the model than static visual features alone.  This also reduces the need for extensive manual data processing typically required to filter out seasonal variations.

According to figure \ref{fig:timeseries}, it shows consistent expansion of all four glacier lakes from 2014 to 2025. This growth across the region confirms that these lakes are responding to accelerated glacier retreat caused by rising temperature. Seasonal noise, like temporary winter ice, was successfully filtered out because of our temporal-first approach, which highlights the true climatic trend of expansion.

Class imbalance is one of the challenges in glacial lake segmentation, where water pixels typically represent less than 1\% of the total study area.  In such a case, standard pixel accuracy can be misleading, as the model could achieve more than 99\% accuracy by classifying all pixels as background. So, we use the F1 score and Intersection over Union (IoU) to provide more rigorous evaluation. Our model achieved an F1 score of 0.9538, which demonstrates that the pipeline effectively balances precision and recall.

The temporal-first approach is the reason behind the high F1 score and IoU. By analysing the dielectric properties of a pixel over a multi-year time series, the model "learns" to ignore transient spatial noise that typically causes class confusion. Whereas spatial-only models struggle to differentiate deep water and permanent shadow, which leads to poor precision.

\section{Limitations and Technical Challenges}
 Despite the high performance of the temporal-first pipeline, there remain several challenges. Topographic distortion, specifically radar layover and shadowing, is one of the most significant limitations. In the high-relief terrain of the Himalayas, steep slopes can obstruct the radar signal, which creates blind spots where lake boundaries cannot be accurately mapped. Furthermore, the spatial resolution of Sentinel-1 (20 m) introduces a “discretisation error” at the lake-land interface. Subtle changes in shoreline position that occur at a sub-pixel level remain undetectable, which may slightly underestimate or overestimate the area of smaller glacial ponds.

 Another demerit involves seasonal ambiguity during winter months. When a lake freezes during the winter season, its dielectric properties shift, which causes the backscatter signature to resemble the surrounding permafrost.

 Our temporal approach mitigates this problem by focusing on summer-maxima trends, but the model’s reliability decreases during peak winter, which may require future integration of optical sensors to resolve these problems. Finally, processing a decade-long time series data is computationally heavier than single-epoch methods. Scaling this pipeline will require optimised horizontal scaling and high-performance computing resources to manage the massive influx of multi-temporal SAR data.

The proposed operational pipeline extracts images from the OPERA RTC burst dataset with 30M spatial resolution. While 30M resolution can capture details up to $900m^2$, features with smaller dimensions aren’t captured. Although the change in area is expected to be higher than $900m^2$ during or before a GLOF event, a better spatial resolution (10 m or 20 m) would be able to capture 2.25 to 9 times smaller day-to-day changes. This requires custom processing of S1 burst raw images, which is out of the current scope of the study. Moreover, the current open access SAR data acquisition source is limited to Sentinel-1. Thus, future work involves incorporating multispectral data sources following the proposed general architecture. With the combined data sources such as Sentinel-1, Sentinel-2, NISAR, Landsat, etc., and retraining of the U-Net model, we can expect a mean revisit time of 1-2 days. This would enable near real-time monitoring of high-risk glacial lakes for GLOF EWS.

\section{Conclusion}
We presented a temporal-first approach to semantic segmentation of Himalayan glacial lakes, achieving an IoU of 0.9130 on high-risk targets. By restricting the training domain to specific lake cohorts, we mitigated common SAR limitations such as topographic noise. Furthermore, the development of a containerized, automated pipeline demonstrates the feasibility of Near Real-Time GLOF monitoring, providing a scalable architectural foundation.

\section*{Acknowledgments}
The authors acknowledge the Space Research Center at the Nepal Academy of Science and Technology (NAST) for providing the research environment. We thank the open-source community for the Sentinel-1 data via the ASF API and NASA HyP3 pipeline.

\bibliographystyle{ieeetr}
\bibliography{bib}

\end{document}